\documentclass[a4paper,12pt]{article}

\usepackage[T2A]{fontenc}
\usepackage{amsmath,amssymb}
\usepackage{cite}
\usepackage[linktocpage=true,plainpages=false,pdfpagelabels=false]{hyperref}

\usepackage[utf8]{inputenc}
\usepackage[english]{babel}

\usepackage{pgfplots}

\newcommand{\dd}{\partial}
\newcommand{\de}{\delta}
\newcommand{\m}{\mu}
\newcommand{\n}{\nu}
\newcommand{\ls}{\left(}
\newcommand{\rs}{\right)}
\newcommand{\al}{\alpha}
\newcommand{\ff}{\varphi}
\newcommand{\ta}{\tau}
\newcommand{\ti}{\tilde}
\newcommand{\dz}{\zeta}
\newcommand{\str}[1]{\mathrel{\mathop{\longrightarrow}\limits_{#1}}}
\newcommand{\ka}{\varkappa}
\newcommand{\be}{\beta}
\newcommand{\ga}{\gamma}

\newcommand{\disn}[2]{$$\displaylines{\refstepcounter{equation}%
            \label{#1}\hskip 1em minus 1em #2\hfilneg}$$}
\newcommand{\nom}{\hfil\hskip 1em minus 1em (\theequation)}

\begin{document}

\title{Possible types of dark matter condensation\\ in embedding gravity}

\author{
S.~A.~Paston\thanks{E-mail: pastonsergey@gmail.com},
A.~J.~Ziyatdinov\thanks{E-mail: a.2iyat@yandex.ru}\\
{\it Saint Petersburg State University, Saint Petersburg, Russia}
}
\date{\vskip 15mm}
\maketitle

\begin{abstract}
We investigate the possibility of explaining the observed effects usually attributed to the existence of dark matter
through a transition from GR to a modified theory of gravity -- embedding gravity.
Since this theory can be reformulated as GR with additional fictitious matter of embedding gravity (FMEG),
which moves independently of ordinary matter, we analyse solutions in which FMEG behaves similarly to cold dark matter.
An upper bound on the possible density of FMEG is obtained, which explains the absence of dark matter effects on small scales.
Possible static condensed structures of FMEG are found, which can be reduced to configurations of the types wall, string, and sphere.
In the latter case, FMEG exhibits the properties of an isothermal ideal gas which has a linear equation of state.
The emerging spherical condensations of FMEG create potential wells that facilitate galaxy formation.
For large values of the radius, the corresponding density distribution profile behaves in the same way as the pseudo-isothermal profile (ISO),
which is successfully employed in fitting galactic dark halo regions, and provides flat galactic rotation curves.
\end{abstract}

\newpage

\section{Introduction}\label{vved}
The puzzle of the nature of dark matter (DM) remains one of the most intriguing problems of contemporary theoretical physics.
Numerous ideas have been proposed concerning the possible nature of dark matter
\cite{1605.04909,DEL_POPOLO_2014,ufn-zasov,2007.15539,2111.00363},
but attempts to detect it directly have so far been unsuccessful \cite{1602.03781,1604.00014}.
There are also approaches in which the effects attributed to DM are explained by some gravity modification
\cite{odintsov2011,1108.6266,1605.04909,2007.15539}.
Such approaches can be divided into those in which DM possesses its own
dynamical degrees of freedom, and therefore can move independently
of ordinary matter, and those in which such degrees of freedom are absent --
the most prominent example being the MOND approach \cite{mond}.
Among the former, one may mention mimetic gravity \cite{mukhanov},
which has been extensively studied over the past decade \cite{mimetic-review17}.
In its original form, this theory yields DM as a dust-like medium with potential motion;
however, various generalisations exist in which the laws governing the behaviour of DM
become more intricate.

In this work, we present results concerning the behaviour of DM
when embedding gravity (also referred to as embedding theory or Regge-Teitelboim gravity) is employed as a modified theory of gravity,
originally proposed 50 years ago \cite{regge}. Initially, this theory was considered as a more
geometric (\emph{a la string}) reformulation of GR, intended
to provide a more convenient framework for quantization.
However, it was observed that this theory differs from GR, its equations of motion take the form \disn{n1}{ D_\m\bigl(\ls G^{\m\n}-\ka\, T^{\m\n}\rs\dd_\n y^a\bigr)=0, \nom} where $y^a$ is the embedding function, $D_\m$ denotes the covariant derivative, $G^{\m\n}$ is the Einstein tensor and $T^{\m\n}$ is the energy-momentum tensor of ordinary matter. It is easy to see that all solutions of the Einstein equations are solutions of these equations as well, but additional solutions also exist.

It was noticed long ago that the equations of motion \eqref{n1}
can be rewritten as the Einstein equations with an additional fictitious matter term \cite{pavsic85}, in which one may attempt to recognise DM:
 \disn{v1}{
G^{\m\n}=\ka \ls T^{\m\n}+\ta^{\m\n}\rs,
\nom}
\vskip -2em
 \disn{v2}{
D_\m\Bigl(\ta^{\m\n}\dd_\n y^a\Bigr)=0.
\nom}
Within this approach
the quantities $y^a$ and $\ta^{\m\n}=\ta^{\n\m}$ describe the fictitious matter of embedding gravity (FMEG),
with $\ta^{\m\n}$ playing the role of its energy-momentum tensor.
Equation \eqref{v1} is the Einstein equation,
while equation \eqref{v2} constrains the behaviour of the fictitious matter.

It should be noted that embedding gravity, like mimetic gravity, arises from a differential
change of variables \cite{statja60} in the Einstein-Hilbert action.
Unlike mimetic gravity, however, where the substitution is rather artificial in form,
the substitution leading to embedding gravity has a clear geometric meaning --
it is the formula for the induced metric of a four-dimensional surface
in the flat ambient space $R^{1,9}$, defined by the embedding function $y^a(x^\m)$
(here and below $\m,\n,...=0,..,3$; $a,b,...=0,..,9$):
\disn{v3}{
g_{\m\n}=(\dd_\m y^a)(\dd_\n y^b)\,\eta_{ab}.
\nom}
Thus, the embedding gravity approach is based on a simple assumption about the nature of our
four-dimensional spacetime -- namely, that it is not an abstract pseudo-Riemannian manifold,
but rather a surface in a flat space of higher dimension.

The system of equations \eqref{v1}-\eqref{v3} is fully equivalent to the equations of motion
of embedding gravity. If necessary, one may also write the action that leads to these equations
in the form of the GR action with an additional contribution corresponding to FMEG,
which is described by the independent variables $y^a$ and $\ta^{\m\n}$ \cite{statja48}.
Rewriting equation \eqref{v2} in the form
\disn{n2}{
\ls D_\m \ta^{\m\n}\rs \dd_\n y^a+\ta^{\m\n}D_\m\dd_\n y^a=0
\nom}
and noting that, with respect to the free index $a$, the first term is tangent to the surface,
whereas the second term is orthogonal to it (as follows from the well-known properties
of the second fundamental form $b^a_{\m\n}=D_\m\dd_\n y^a$ of the surface),
equation \eqref{v2} can be rewritten as a system of equations
\disn{v4}{
D_\m\ta^{\m\n}=0,
\nom}\vskip -2em
\disn{v5}{
\ta^{\m\n}b^a_{\m\n}=0,
\nom}
see details, e.g., in \cite{statja68}.
Equation \eqref{v4} plays the role of the standard momentum constraint for the fictitious matter.

The presence or absence of FMEG in the Universe is determined by the initial data
emerging in the early Universe. It was shown in \cite{statja18} that it is sufficient,
at some initial moment in time, to require the condition $\ta^{0\n}=0$
for the usual Einstein equations to hold from that moment on, i.e. for $\ta^{\m\n}=0$
and for FMEG to be absent. In the present work we shall assume that the initial data
are such that FMEG is present, i.e. $\ta^{\m\n}\ne0$.

We attempt to identify FMEG with DM (assuming it to be cold DM),
studying its behaviour on scales smaller than cosmological ones,
at sufficiently late stages of cosmic expansion,
when the matter density is already not very high.
Then, neglecting small regions in the vicinity of black holes and neutron stars,
we may consider the gravitational field to be weak:
\disn{k4}{
g_{\m\n}=\eta_{\m\n}+h_{\m\n},
\nom}
where $\eta_{\m\n}$ is the flat metric in Lorentzian coordinates, and $h_{\m\n}\ll 1$.
On the other hand, we are interested in solutions of the system of equations
\eqref{v1},\eqref{v3},\eqref{v4},\eqref{v5}, for which the energy-momentum tensor of FMEG $\ta^{\m\n}$
corresponds to non-relativistic matter.
This means that, in the leading approximation, it has a single non-zero component $\ta^{00}$,
i.e. $\ta^{\m\n}=\rho_\ta\de^\m_0\de^\n_0$.
Neglecting the effects of curved spacetime (in particular, when we consider not too long time intervals during which no significant cosmic expansion occurs),
equation \eqref{v4} reduces approximately to $\dd_\m\ta^{\m\n}=0$, i.e. to local conservation of FMEG energy and momentum.
In the leading approximation this reduces to the condition $\dd_0 \rho_\ta=0$, i.e. constancy of its density in time,
while the spatial distribution of $\rho_\ta$ is not constrained by the equations at this order.

As usual, we assume that at present, on cosmological scales,
three-dimensional homogeneity and isotropy are valid for both the metric and the distribution of ordinary matter
(while no analogous symmetry requirement is imposed on the quantity $y^a$).
Consequently, according to equation \eqref{v1}, the FMEG density $\rho_\ta$, when averaged
over cosmological scales, takes a constant value.
The sign of this value may, in principle, be arbitrary -- it depends on the preceding history,
which led to the formation of a three-dimensionally homogeneous and isotropic configuration.
The analysis of such a formation, typically associated with inflationary expansion,
lies beyond the scope of the present work.
It is usually assumed that DM creates potential wells that attract ordinary matter,
thus assisting galaxy formation.
This corresponds to $\rho_\ta>0$, and we therefore restrict ourselves to consideration of this case.
We assume that initially $\rho_\ta$ was almost independent of the spatial coordinates
(e.g. fluctuations not exceeding $10^{-5}$ at the time of recombination),
and later static condensations began to form from these fluctuations.
Determination possible configurations of such condensations is the main goal of this work.

Since the metric $g_{\m\n}$ is expressed via \eqref{v3} in terms of the embedding function $y^a$,
writing the metric in the form \eqref{k4} means that the surface defined by $y^a$
may be obtained as a small deformation of a surface with flat induced metric:
\disn{k5}{
y^a= \bar y^a+q^a,
\nom}
where $\bar y^a$ describes some background surface with metric $\eta_{\m\n}$, and $q^a$ is small.
From \eqref{v3}, in the linear approximation, the relation between the metric deformation and the embedding function follows \cite{statja77}:
\disn{k20}{
h_{\m\n}=\dd_\m\xi_\n+\dd_\n\xi_\m-2q_a \bar b^a_{\m\n},
\nom}
where $\xi_\m=q_a\dd_\m \bar y^a$, and $\bar b^a_{\m\n}=\dd_\m\dd_\n \bar y^a$ is the second fundamental form of the background surface.
Here $\xi_\m$ contains only the components of the deformation $q^a$ tangent to the background surface,
while the last term contains only the transverse ones, since the second fundamental form possesses the property of transversality:
$\dz_a b^a_{\m\n}=0$ for any vector $\dz_a$ tangent to the surface.

In solving the system of equations \eqref{v1},\eqref{v3},\eqref{v4},\eqref{v5},
the background $\bar y^a$ may be taken as any embedding of the Minkowski space metric.
Therefore, in the choice of a specific background there exists a substantial arbitrariness \cite{statja71},
equivalent to specifying initial data.
If $\bar y^a$ is taken as the simplest embedding corresponding to the four-dimensional plane,
then $\bar b^a_{\m\n}=0$, and there will be no linear relation between $h_{\m\n}$ and $q_a$ (each containing 10 independent components).
As a result, above such a background the equations do not linearise, and the principle of superposition
for weak gravitational fields does not hold \cite{statja77} (see also the discussion of this issue below, in section \ref{linreg}).
To avoid this, all 10 components of $q_a$ must be expressed linearly in terms of $h_{\m\n}$,
and therefore we shall assume that $\bar y^a$ is an \emph{unfolded embedding} \cite{statja71}
(note that this corresponds to the generic case).
This means that $\bar b^a_{\m\n}$ has the maximal possible rank 6,
when regarded as a $10\times 10$ matrix with the pair of symmetric indices $\m,\n$ treated as a multi-index taking 10 values.
The rank 6 turns out to be maximal owing to the above-mentioned transversality property of the second fundamental form.

If $q^a$ is sufficiently small (together with its derivatives), i.e. the gravitational field is sufficiently weak, then
\disn{k13}{
b^a_{\m\n}\approx\bar b^a_{\m\n}.
\nom}
If in this case we are interested in solutions with the structure of $\tau^{\m\n}$ described below \eqref{k4},
then in view of \eqref{v5} it must hold that $\dd_0^2\bar y^a\ll \dd_i\dd_k\bar y^a$ (here and below $i,k,...=1,2,3$).
We shall further assume that the maximal possible rank 6 is possessed not only by $\bar b^a_{\m\n}$,
but also by $\bar b^a_{ik}$ (in the same sense), which means that the submanifold $x^0=\mathrm{const}$ of the background surface is unfolded.
If this assumption is not made, then FMEG will not remain non-relativistic (see section \ref{linreg}).
The described choice of background embedding function corresponds (at least when considering
a not too large spacetime region) to that used in \cite{statja68},
where the non-relativistic limit of embedding gravity was studied.
There, non-relativistic equations of motion for FMEG were obtained,
but these dynamical equations turned out to be too complicated to aim for analytic solutions.

In the present work, we shall seek the solutions describing possible
static configurations of FMEG
(in fact, we find configurations that are static only in the leading approximation,
and therefore, strictly speaking, they should be referred to as \textit{approximately static} configurations),
i.e. possible types of condensation of this matter
that may arise from an initially almost homogeneous distribution.
To simplify the task, we consider such a spacetime region
in which the quantity $\bar b^a_{\m\n}$ varies little, remaining approximately constant.
We assume that the size of this region is sufficiently large; a specific estimate will be made below, see the end of section~\ref{linreg}.
For a given value of $\bar b^a_{\m\n}$, Lorentzian coordinates in the ambient space can be chosen such that the values $a=0,..,3$
correspond to directions tangent to the surface.
We then introduce the index $A$ running over the remaining values $4,..,9$.
In the region under consideration we may then write
\disn{k11}{
\bar b^a_{\m\n}=\de^a_A B^A_{\m\n},
\nom}
where $B^A_{\m\n}=B^A_{\n\m}$ is a set of constants of dimension inverse length.
We denote by $L$ the characteristic length associated with the largest component of $B^A_{\m\n}$,
so that $B^A_{\m\n}\sim 1/L$ for at least some of its components.
The quantity $L$ has the geometric meaning of the outer curvature radius of the surface.
At the same time, its intrinsic curvature vanishes (i.e. the Riemann curvature tensor is zero),
since the background surface $\bar y^a$ has a flat metric.
According to the Gauss equation \cite{kobno2}, this imposes the constraint
\disn{k12}{
B^A_{\m\al}B^A_{\n\be}-B^A_{\m\be}B^A_{\n\al}=0.
\nom}
The size of the spacetime region in which $\bar b^a_{\m\n}$ remains constant
and in which we shall analyse the solutions must be small compared with $L$.

In section~\ref{linreg} we analyse the equations of motion of the theory and find that,
within a finite region and under sufficiently weak gravity,
the equation governing the behaviour of FMEG becomes linear.
As gravity strengthens, the linear regime breaks down,
leading to a restriction on the maximum density of FMEG for its static configurations.
In section~\ref{tipi} we classify and begin to study possible static condensations of FMEG,
which may arise in the region of validity of the linear regime in the absence of ordinary matter.
The most interesting case, that of spherical condensation, is investigated in section~\ref{sfer}.
In section~\ref{zakl} we discuss the results obtained.

\section{Linear regime of FMEG and its maximum condensation}\label{linreg}
We will analyse the system of equations \eqref{v1},\eqref{v3},\eqref{v4},\eqref{v5} in a finite region of spacetime,
in which the quantity $\bar b^a_{\m\n}$ remains constant and is given by \eqref{k11}.
We neglect the contribution of gravitational waves to the correction $h_{\m\n}$ of the metric,
which is negligibly small at present.
Then $h_{\m\n}$ (and consequently the deformation $q^a$ from \eqref{k5},
which is expressed linearly in terms of $h_{\m\n}$ under the assumption of the unfolded background $\bar y^a$)
is determined by the right-hand side $T^{\m\n}+\ta^{\m\n}$
of the Einstein equation \eqref{v1}, which can be linearised in the standard way.
This means that the presence of matter (both ordinary and FMEG) leads to a deviation
of the surface corresponding to $y^a$ from the background one corresponding to $\bar y^a$.
If $T^{\m\n}+\ta^{\m\n}$ is sufficiently small, this leads to $q^a$ being sufficiently small
for \eqref{k13} to hold.
Then equation \eqref{v5} in the region under consideration, in the leading approximation, takes the form
\disn{k14}{
\ta^{\m\n}B^A_{\m\n}=0,
\nom}
and we shall say that FMEG is in the \emph{linear regime},
since $B^A_{\m\n}$ is constant and this equation is linear in $\ta^{\m\n}$.
In the general case, when \eqref{k13} does not hold,
the influence of $\ta^{\m\n}$ on $q^a$ through $h_{\m\n}$ affects $b^a_{\m\n}$,
and equation \eqref{v5} becomes nonlinear in $\ta^{\m\n}$.
From the geometric point of view, the validity of \eqref{k13} corresponds to the situation
in which the intrinsic curvature (described by the Riemann curvature tensor)
of the surface representing our spacetime is much smaller
than its extrinsic curvature (described by the second fundamental form of the surface).

As mentioned in the Introduction, we assume that $\bar b^a_{ik}$, and hence $B^A_{ik}$,
has the maximal possible rank 6.
If this assumption is not made, then equation \eqref{k14} admits solutions with arbitrarily large nonzero values
of the stress tensor $\ta^{ik}$, which according to equation \eqref{v4} would lead to non-relativistic behaviour of FMEG.
Such a rank of $B^A_{ik}$ allows one to introduce a reciprocal quantity
$\al^{ik}_A$, uniquely defined by the relations:
\disn{k15}{
\al^{ik}_A=\al^{ki}_A,\qquad
\al^{ik}_A B_{ik}^B=\de^B_A.
\nom}
We assume that all eigenvalues of the $6\times 6$ matrix $B^A_{ik}$ (treating the pair of symmetric indices $i,k$
as a multi-index running over six values) are of the same order $1/L$.
Then the eigenvalues of $\al^{ik}_A$ will be of order $L$.
Using the introduced quantity $\al^{ik}_A$, equation \eqref{k14} in the linear regime of FMEG
can be written as
\disn{k16}{
\ta^{ik}=-\al^{ik}_A\ls \ta^{00}B_{00}^A+2\ta^{0m}B_{0m}^A\rs.
\nom}
As already mentioned above, in order for the properties of FMEG to resemble the observed properties of cold DM,
we restrict our study to solutions for which $\ta^{0m},\ta^{ik}\ll\ta^{00}\equiv\rho_\ta$.
For the same reason, we assume that $\rho_\ta\ge0$.
In addition, in this work we restrict ourselves to solutions for which the second term
on the right-hand side of equation \eqref{k16} can be neglected compared with the first.
Given $\ta^{0m}\ll\ta^{00}$, it suffices for this that $B_{0m}^A$ is not much larger than $B_{00}^A$.
The validity of all the assumptions made should follow from a suitable choice of initial data
when solving the system of equations \eqref{v1},\eqref{v3},\eqref{v4},\eqref{v5}.

As a result of the above assumption, relation \eqref{k16} can be written as
\disn{k34}{
\ta^{ik}= w^{ik}\rho_\ta,
\nom}\vskip -2em
\disn{k34.2}{
w^{ik}=-B_{00}^A\al^{ik}_A.
\nom}
Here the quantity $w^{ik}$ is approximately a set of constants in the region of spacetime considered,
while in general it is a slowly varying function.
It varies significantly on scales comparable with the quantity $L$
(the characteristic scale for $\bar b^a_{\m\n}$),
which itself also changes slowly on these scales.
As a result, in the linear regime FMEG possesses a stress tensor
proportional to the matter density, with a slowly varying matrix coefficient.
The assumption $\ta^{ik}\ll\rho_\ta$ corresponds to the condition $w^{ik}\ll 1$,
from which it follows, by \eqref{k15}, that $B_{00}^A\ll 1/L$.

As long as the linear regime is preserved, FMEG can condense, forming an equilibrium (i.e., approximately static) configuration in which gravitational forces are balanced by pressure (which, generally speaking, can be anisotropic). However, as a result of condensation, the FMEG density $\rho_\tau$ may increase to the point where $T^{\mu\nu}+\tau^{\mu\nu}$ is no longer sufficiently small, and FMEG leaves the linear regime since \eqref{k13} no longer holds. Consequently, the stress tensor $\tau^{ik}$ will no longer be determined by the linear equation \eqref{k14}; instead, the more general nonlinear equation \eqref{v5} comes into play.
Using \eqref{k5}, this equation can be rewritten as
\disn{n3}{
\ta^{\m\n}D_\m\dd_\n\ls \bar y^a+q^a\rs=0,
\nom}
where $q^a$ is expressed in terms of the metric perturbation $h_{\m\n}$ as the solution of equation \eqref{v3}
(note that outside the linear regime equation \eqref{k20} can no longer be used).
Repeating the reasoning given above and again using the constraint $\ta^{0m}\ll\ta^{00}$, from equation \eqref{n3}
one can once again obtain formula \eqref{k34}, but now the quantity $w^{ik}$ will no longer be a set of constants,
and will instead depend significantly on $q^a$, and hence will be some nonlocal function
(since $q^a$ is obtained as a solution of the partial differential equation \eqref{v3}) of $h_{\m\n}$.

From the equation of motion \eqref{v4} in the leading approximation one can derive the dynamical equation
\disn{n4}{
\dd_0\ta^{0k}=-\rho_{\ta}\dd_k\ff-\dd_i\ls w^{ik}\rho_{\ta}\rs,
\nom}
where on the right-hand side the first term corresponds to the ordinary gravitational force,
while the second term defines the \textit{self-interaction force} associated with the presence of a certain anisotropic pressure of FMEG.
Here and below $\varphi$ is the Newtonian gravitational potential, we use units with $c=1$ and signature $(+-\ldots-)$.
A necessary condition for the emergence of a static configuration of FMEG is the time independence of the quantity $w^{ik}$
that influences the self-interaction force:
\disn{n6}{
\dd_0 w^{ik}=0.
\nom}
The quantity $w^{ik}$, according to the nonlinear analogue of equation \eqref{k34.2}, is expressed in terms of $q^a$,
so that condition \eqref{n6} requires the fulfilment of six additional equations for the ten components of $q^a$
in addition to the ten equations \eqref{v3}.
It is important to note that the deformation of the embedding function $q^a$ cannot be completely time-independent,
since in this case, according to \eqref{v3}, the gravitational potential $\ff=h_{00}/2$ would vanish.
Since one obtains sixteen equations for ten unknowns, this provides some grounds for concluding that such a system
has no solutions, and therefore static condensations of FMEG cannot arise outside the linear regime.
A more rigorous proof of this statement at the analytical level is, unfortunately, not feasible
due to the need to analyse nonlinear partial differential equations.
One cannot completely rule out the possibility that this statement is valid only in the generic case,
and that for certain special background embeddings $\bar y^a$
(recall that according to \eqref{k5} $y^a$ is close to $\bar y^a$ since $q^a$ is small) it may fail.
A definitive verification could be attempted through numerical analysis, but this task lies beyond the scope of the present work.

One may say that outside the linear regime, owing to the nonlocal dependence of $w^{ik}$ on the gravitational potential $\ff$,
the self-interaction force has a rather random direction, moreover a time-dependent one. As a result it
can no longer compensate for gravitational forces, and an equilibrium configuration cannot form,
since the self-interaction force ``smears out'' them.
In the linear regime, however, the main contribution to $w^{ik}$ is determined by the background embedding
and is a set of constants \eqref{k34.2}, so in the leading approximation compensation is possible.
Corrections to this leading contribution do depend on time, but they are increasingly suppressed
the better the linear regime holds (and hence the smaller the quantity $T^{\m\n}+\ta^{\m\n}$),
and they can be neglected.
Thus, approximately static configurations of FMEG
(meaning staticity in the sense of neglecting deviations from the linear regime)
can only arise at condensations that are not too large.
This property of FMEG is very important when we attempt to see it as DM. If FMEG could condense arbitrarily strongly, the question would arise: why, besides condensations corresponding to galaxies, do we not observe much denser condensations comparable in density and scale to stars or planets.

Let us estimate the boundary value $\rho_b$ of the total density $\rho+\rho_\tau$ (here $\rho\equiv T^{00}$ is the density of ordinary matter, assumed to be nonrelativistic), at which FMEG leaves the linear regime. For static distributions of ordinary matter and FMEG the condition $\rho+\rho_\tau<\rho_b$ must hold, which also implies $\rho_\tau<\rho_b$. Using \eqref{k11} and additionally choosing Lorentz coordinates in the embedding space so that approximately $\partial_\mu \bar y^a=\delta^a_\mu$ (which can be done in the spacetime region under consideration, small compared to $L$), formula \eqref{k20} can be rewritten as
\disn{k21}{
h_{\mu\nu}=\partial_\mu q_\nu+\partial_\nu q_\mu-2q_A B^A_{\mu\nu}.
\nom}
On the other hand, since all matter is nonrelativistic, the solution of the linearized Einstein equations in harmonic coordinates yields
\disn{k23}{
h_{\mu\nu}=2\varphi\delta_{\mu\nu}.
\nom}

The exact expression for $b^a_{\mu\nu}$ is
\disn{k26}{
b^a_{\mu\nu}
=\partial_\mu\partial_\nu y^a-\Gamma^\alpha_{\mu\nu}\partial_\alpha y^a=
\bar b^a_{\mu\nu}+\partial_\mu\partial_\nu q^a-\Gamma^\alpha_{\mu\nu}\partial_\alpha y^a.
\nom}
Since $B_{\mu\nu}^A\sim 1/L$, from \eqref{k21} it follows that $q_A\sim L\varphi$ (taking into account that the terms with $q_\mu$ are coordinate-dependent and do not contribute to curvature). Roughly assuming that under differentiation of $\varphi$ we have $\partial_k\sim 1/r$, $\partial_0\sim 1/t$, where $r$ and $t$ are the characteristic spatial and temporal scales of variation of $\varphi$ (with $t\gg r$ in nonrelativistic motion), from \eqref{k26} we obtain an estimate for the transverse (in index $a$) component of the deformation of the second fundamental form:
\disn{k27}{
b^A_{ik}-\bar b^A_{ik}\sim\frac{L}{r^2}\varphi,\qquad
b^A_{00}-\bar b^A_{00}\sim\frac{L}{t^2}\varphi.
\nom}
The longitudinal part can be estimated as $\varphi/r$, but it does not contribute to $w^{ik}$. Since $\bar b^a_{ik}\sim 1/L$ and $\bar b^a_{00}\sim w/L$ (where $w\ll 1$ is the characteristic value of $w^{ik}$, see \eqref{k34.2}), the condition \eqref{k13} for the validity of the linear regime of FMEG can be written as simultaneous fulfillment of $\varphi/r^2\ll 1/L^2$ and $\varphi/t^2\ll w/L^2$.
When studying static configurations of fictitious matter, the latter condition is automatically satisfied since in this case $t\to\infty$. Thus, it may only become relevant when studying relaxation processes. For the self-consistency of static configurations, only the condition $\varphi/r^2\ll 1/L^2$ is important.
If we again take into account the earlier estimate $\partial_k\sim 1/r$ acting on $\varphi$, and the fact that $\varphi$ satisfies the equation $\partial^2\varphi=-4\pi G(\rho+\rho_\tau)$ (where $G$ is Newton's gravitational constant), we obtain the boundary value $\rho_b$ for the total matter density:
\disn{k30}{
\rho+\rho_\tau\ll\rho_b,\qquad
\rho_b=\frac{1}{4\pi GL^2}.
\nom}

If this condition is satisfied and the gravitational field does not vary too rapidly in time, then FMEG remains in the linear regime. In this case, it is easy to see that the principle of superposition holds for the weak gravitational field generated by ordinary matter \cite{statja77}. Indeed, after linearization of the Einstein equations \eqref{v1}, the contribution to $h_{\mu\nu}$ corresponding to the density of ordinary matter $\rho$ is linear in $\rho$, while the contribution from a fixed value of $\rho_\tau$ remains unchanged. The possibility of keeping $\rho_\tau$ fixed is ensured by the independence of $\tau^{\mu\nu}$ from $h_{\mu\nu}$ due to the validity of equation \eqref{k14} in the linear regime of FMEG (note the importance of using an unfolded embedding as the background, without which the linear regime of FMEG is impossible).
When the total density $\rho+\rho_\tau$ increases such that the linear regime is violated and $\rho$ and $\rho_\tau$ become comparable in magnitude, the principle of superposition for gravitating ordinary matter no longer holds, since instead of \eqref{k14} the original equation \eqref{v5} applies, and $\tau^{\mu\nu}$ begins to depend on $h_{\mu\nu}$.

It is important to note that in the case $\rho_\tau\ll\rho$ (which holds in particular if $\rho\gg\rho_b$), according to \eqref{v1}, the effect of FMEG on the correction $h_{\mu\nu}$ to the metric becomes negligible compared to the effect of ordinary matter. In this case, the principle of superposition for the weak gravitational field generated by ordinary matter is again valid. This occurs despite the fact that FMEG is not in the linear regime -- simply because although the energy-momentum tensor $\tau^{\mu\nu}$ changes with $\rho$ in a complicated way, it does not affect the metric due to its smallness, while what is observable is precisely the metric.
As a result, at scales where $\rho\gg\rho_b$, the influence of FMEG on the gravitational field becomes practically unnoticeable. Therefore, for a suitable value of $\rho_b$, the influence of FMEG on the gravitational potential on scales such as the Solar System turns out to be negligible.

One can note a certain similarity, but also a difference, between the discussed approach and the MOND paradigm \cite{mond}. In MOND, modification of the law generating the gravitational field occurs when a boundary value of the acceleration $\partial_k\varphi\sim\varphi/r$ is reached. In the present approach, such a modification, being a consequence of the significant influence of fictitious matter, occurs when the boundary value of the second derivative $\partial_i\partial_k\varphi\sim\varphi/r^2$ is reached. The contribution of FMEG to this quantity, according to the Poisson equation, is bounded by $4\pi G\rho_\tau$, i.e., $4\pi G\rho_b$, and when $\partial_i\partial_k\varphi$ decreases to this boundary value, fictitious matter begins to noticeably influence the gravitational field.

The value $L$, which sets the scale of $\bar b^a_{\mu\nu}$, is determined by the behavior of the background embedding $y^a$ in the spacetime region under consideration. In finding static configurations of FMEG in this region, it serves as a free parameter characterizing the background surface at present. According to \eqref{k30}, it sets the maximum density to which FMEG can condense.
To estimate the possible value of $L$ from observations for the region of space containing our galaxy, one can take the value of the total matter density $0.1\,M_{\odot}/\text{pc}^3$, known for the Solar neighborhood on a scale of $1\text{kpc}$ \cite{GalDyn,ufn-zasov}. Since there are reasons to believe that at these scales the influence of DM is already noticeable (the DM density is estimated at $0.01\,M_{\odot}/\text{pc}^3$, which, however, is close to the uncertainty \cite{GalDyn,ufn-zasov}), one can assume that FMEG is already in the linear regime here. Then one can set $\rho_b=0.1\,M_{\odot}/\text{pc}^3$, which by \eqref{k30} gives
\disn{k31}{
L=4\,\text{Mpc}.
\nom}
It may turn out that the linear regime for such density has not yet been established, since for these scales it is difficult to obtain data on the static distribution of DM. Therefore, the value of the parameter $L$ for the vicinity of our galaxy may be larger than the value in \eqref{k31}.
On the other hand, new data from pulsar timing \cite{2507.16932} may point in favour of higher values of $\rho_b$ as well --
the authors discuss a possible density value up to $\rho_b=10\,M_{\odot}/\text{pc}^3$. It is not excluded that this estimate
will be refined in the future, since the density profile of the sub-halo cannot currently be determined.
If one proceeds from such a (possibly overestimated) value, then the value of $L$ must be reduced by roughly a factor of 10
compared with \eqref{k31}.
It should be noted that in any case, the size of the region of space under consideration, within which $\bar b^a_{\mu\nu}$ and hence $w^{ik}$ vary little (recall that this size must be smaller than $L$), turns out to be quite large -- comparable to the size of the Local Group of galaxies.

The theoretical determination of possible values of $L$, as well as of other characteristics of the background surface at present, requires a detailed study of the process of cosmic expansion within the framework of embedding theory. Such studies have so far been carried out only for a 5-dimensional embedding space \cite{davids01,statja26}. Therefore, the task arises of conducting a similar study for unfolded embeddings into a 10-dimensional ambient space. Such an investigation lies beyond the scope of the present work.

\section{Types of static condensations}\label{tipi}
Let us examine which static distributions of FMEG may arise when it is in the linear regime and ordinary matter is absent.
We assume that they originate from an initially almost homogeneous distribution with
$\ta^{\m\n}=\rho_\ta\de^\m_0\de^\n_0$, $\rho_\ta>0$, as a result of random fluctuations.
From equation \eqref{v4}, assuming the fields are static and $w_{ik}$ in the region under study is constant,
and taking into account \eqref{k34}, \eqref{k23}, in the leading approximation we obtain the equation
\disn{k39}{
\rho_\ta\dd_i\ff+w^{ik}\dd_k\rho_\ta=0.
\nom}
If $\rho_\ta\ne0$ (a zero density of FMEG yields a trivial solution), this equation can be rewritten as
\disn{k40}{
\dd_i\ff+w^{ik}\dd_k z=0,
\nom}
where $z=\log(\rho_\ta/\tilde\rho)$, and $\tilde\rho$ is an arbitrary positive dimensionless constant.
By rotating the coordinate axes, the matrix $w^{ik}$ can be diagonalized:
$w^{ik}=diag\{w_1,w_2,w_3\}$, where $w_j$ are the eigenvalues.
Applying differentiation $\dd_m$ to \eqref{k40} and antisymmetrizing with respect to $i,m$, we obtain
\disn{k42}{
(w_1-w_2)\dd_1\dd_2 z=0,\quad
(w_1-w_3)\dd_1\dd_3 z=0,\quad
(w_2-w_3)\dd_2\dd_3 z=0,\quad
\nom}
where it is taken into account that $w^{ik}$ is constant in the region under consideration.
If, instead, one applies differentiation $\dd_m$ to \eqref{k40}, then multiplies by $w^{nm}$,
and afterwards antisymmetrizes with respect to $i,n$, one obtains
\disn{k42.1}{
(w_1-w_2)\dd_1\dd_2 \ff=0,\quad
(w_1-w_3)\dd_1\dd_3 \ff=0,\quad
(w_2-w_3)\dd_2\dd_3 \ff=0.\quad
\nom}

First, let us consider the situation when all $w_j$ are distinct.
In this case, from \eqref{k42} and \eqref{k42.1} we find that both $\ff$ and $z$ must be sums of three contributions,
each of which is a function of only one of the coordinates $x^1$, $x^2$, $x^3$.
Let us take into account that when the ordinary matter is absent and fields are static, in the leading approximation,
the Poisson equation holds:
\disn{k47}{
\dd_i\dd_i\ff=4\pi G \tilde\rho\, e^z,
\nom}
where we have used $\rho_\ta=\tilde\rho e^z$.
Since the left-hand side of this equation turns out to be a sum of contributions, each depending on only one coordinate,
whereas the right-hand side is a product of such contributions,
the equation can be satisfied only if two of these contributions are constants.
As a result, the solution will be only a potential $\ff$ depending on one of the coordinates corresponding to the basis
in which $w^{ik}$ is diagonal; denote this coordinate by $x^1$, and for convenience set $w\equiv w_1$.

The case $w\le0$ corresponds to nonpositive pressure of FMEG and does not lead to the emergence of its static condensations
with density decreasing at infinity, and therefore we shall study the case $w>0$.
Then from \eqref{k40} we find $z=C_1-\ff/w$, where $C_1$ is an arbitrary constant.
Substituting this into \eqref{k47} we obtain the equation
\disn{k50}{
\ff''\ls x^1\rs=C_2 e^{-\ff(x^1)/w},
\nom}
where $C_2=4\pi G\tilde\rho e^{C_1}$ is a new arbitrary constant, which is positive.
The solution of this equation can be written in the form
\disn{k52}{
\ff=\ff_0+2w\log\ls\cosh\frac{x^1-x^1_0}{\de}\rs,
\nom}
where $\ff_0$, $x^1_0$, and $\de$ are the parameters characterising the solution (the arbitrariness of one of them is related
to the arbitrariness of $C_2$). The density distribution of FMEG corresponding to this solution has the form
\disn{k53}{
\rho_\ta=\frac{\rho_0}{\cosh^2\frac{x^1-x^1_0}{\de}},
\nom}
where $\rho_0=w/(2\pi G\de^2)$.
The obtained static condensation of FMEG, corresponding to the case when the matrix $w^{ik}$ has all different eigenvalues,
has the form of a thick wall. This condensation is characterised by two parameters: $x^1_0$ sets the position of the wall centre,
and $\de$ determines its thickness. As can be seen, $\de$ also affects the FMEG density $\rho_0$ at the wall centre.
Along the wall, the density does not vary, but this holds only in the spatial region where $\bar b^a_{\m\n}$
remains approximately constant (see before \eqref{k11}).
Therefore, the longitudinal size of the wall cannot exceed the size of this region, i.e. the quantity $L$.

Next, let us consider the situation when two of the three eigenvalues $w_j$ coincide.
For definiteness, let us assume that $w\equiv w_1=w_2\ne w_3$.
In this case, from \eqref{k42} and \eqref{k42.1} we find that both $\ff$ and $z$ must be sums of two contributions,
one of which depends on $x^1,x^2$, and the other only on $x^3$.
From \eqref{k47} one can again infer that either the first or the second of these contributions must be constants.
The first option gives the case of wall-type condensation already studied above.
Let us consider the second option, when $\ff$ and $\rho_\ta$ depend on $x^1,x^2$.
The case $w\le0$ again does not yield static condensations decreasing at infinity, so we take $w>0$.
Once again, from \eqref{k40} we obtain $z=C_1-\ff/w$, and substituting this into \eqref{k47}, we obtain the equation
\disn{k60}{
\ls\dd_1^2+\dd_2^2\rs\ff=C_2 e^{-\ff/w},
\nom}
where $C_2>0$.
Up to a relabelling, this is the well-known Liouville equation.
All of its solutions, for which the right-hand side of \eqref{k60} (and hence the FMEG density $\rho_\ta$)
decreases sufficiently rapidly at infinity, turn out to be symmetric with respect to rotations in the plane $(x^1,x^2)$
around some centre (which may be taken as the origin) \cite{chen-li}.
For solutions from this class, the equation is rewritten in the form
\disn{k62}{
\ff''(s)+\frac{1}{s}\ff'(s)=C_2 e^{-\ff(s)/w},
\nom}
where $s=\sqrt{{x^1}^2+{x^2}^2}$.

The requirement that the density $\rho_\ta$ have no delta-functional contribution at the centre of symmetry imposes the additional condition $s\ff'(s)\to0$ as $s\to0$.
Solutions of equation \eqref{k62} that satisfy this condition can be written as
\disn{k67}{
\ff(s)=\ff_0+2w\log\ls 1+\frac{s^2}{\de^2}\rs,
\nom}
where $\ff_0$ and $\de$ are again constants parametrising the solution.
The corresponding density distribution of FMEG takes the form
\disn{k69}{
\rho_\ta=\frac{\rho_0}{\ls 1+\dfrac{s^2}{\de^2}\rs^2},
\nom}
with $\rho_0=2w/(\pi G\de^2)$.
Such a static FMEG condensation, corresponding to the case in which the matrix $w^{ik}$ has two coinciding positive eigenvalues, has the form of a thick string.
It is characterised by a single parameter $\de$, which determines the thickness of the string and also affects the central density of FMEG, $\rho_0$.
Similarly to the case of the wall, along the string the density does not change as long as $\bar b^a_{\m\n}$ varies little, and therefore the string length does not exceed $L$.

In addition to the two cases of eigenvalue choices of the matrix $w^{ik}$ considered above, it remains to study the situation in which all three coincide; this is the subject of the next section.
It may be noted that the emergence of different FMEG configurations depending on the values of these eigenvalues bears some resemblance to the appearance in the Zeldovich approach \cite{zeldovich70} of different matter condensations depending on the signs of the eigenvalues of the deformation tensor.

\section{Spherical condensations}\label{sfer}
Let us consider the situation in which all three eigenvalues $w_j$ coincide, so that $w^{ik}=w\de^{ik}$.
Taking into account \eqref{k34}, this means that in this case FMEG behaves like matter with pressure $p_\ta=w \rho_\ta$.
Such an equation of state corresponds to an ideal gas at constant temperature, so that FMEG is an isothermal ideal gas.
From \eqref{k40} we again obtain $z=C_1-\ff/w$. Substituting this into \eqref{k47}, we obtain the equation
\disn{k75}{
\dd_i\dd_i\ff=C_2 e^{-\ff/w},
\nom}
where again $C_2>0$. This is the three-dimensional analogue of the Liouville equation.
As before, the case $w\le0$ does not yield static condensations with density decaying at infinity, so we restrict attention to $w>0$.

Equation \eqref{k75} possesses spherical symmetry, and we confine ourselves to the study of its spherically symmetric solutions.
Since all solutions of this equation correspond to static configurations in which gravitational forces are balanced by pressure forces, it is natural to suppose that, analogous to the two-dimensional case (see the previous section), under the condition of vanishing FMEG density at infinity all solutions should possess spherical symmetry.
However, unlike the two-dimensional case, in three dimensions we do not know a proof of this fact.

For spherically symmetric solutions the equation \eqref{k75} takes the form
\disn{k76}{
\ff''(r)+\frac{2}{r}\ff'(r)=C_2 e^{-\ff(r)/w},
\nom}
where $r=\sqrt{{x^1}^2+{x^2}^2+{x^3}^2}$.
As in the two-dimensional case, in addition to this equation we must require that the density $\rho_\ta$ have no delta-functional contribution at the centre of symmetry, which yields the condition
\disn{k77}{
r^2\ff'(r)\str{r\to0} 0.
\nom}
Solutions of this problem describe the density distribution profile of the self-gravitating isothermal sphere.
This profile arises in the study of stellar dynamics \cite{GalDyn} (Section 4.3.3(b)).
It is not possible to write down all solutions analytically, even using known special functions; however, the solutions can be expressed in terms of a single function, which is easily found numerically and for which all asymptotics are known.

Let us analyse equation \eqref{k76}. Rescaling the independent variable as $r=\ti r\sqrt{w/C_2}$ (so that $\ti r$ is dimensionless) and the unknown function as $\ff=-w\ti\ff$,
we acquire the parameter-free form
\disn{k77.2}{
\ti\ff''(\ti r)+\frac{2}{\ti r}\ti\ff'(\ti r)+e^{\ti\ff(\ti r)}=0.
\nom}
We then set $\ti r=e^u$ and write the desired function as
\disn{k77a}{
\ti\ff(\ti r)=\ga(u)-2u.
\nom}
As a result, for the new unknown function $\ga(u)$ we obtain the equation
\disn{k80}{
\ga''(u)+\ga'(u)-2+e^{\ga(u)}=0.
\nom}
Since the independent variable $u$ does not enter explicitly into this equation, the set of solutions possesses symmetry under shifts in $u$.
This allows the order of the equation to be reduced by introducing $\psi(\ga)=\ga'(u)$, giving
\disn{k82}{
\psi(\ga)\psi'(\ga)+\psi(\ga)-2+e^{\ga}=0.
\nom}

It should be noted that in passing from \eqref{k80} to \eqref{k82}, the constant solution of equation \eqref{k80} is lost, namely
$\ga(u)=\log2$, which corresponds to
\disn{k83}{
\ff(r)=w\log\frac{C_2 r^2}{2w},\qquad
\rho_\ta=\frac{w}{2\pi G\, r^2}.
\nom}
For this solution the condition \eqref{k77} is satisfied. This density profile is known as the \emph{singular isothermal sphere} \cite{GalDyn}.
In this profile the growth of central density is unbounded, so that the condition of the linear regime \eqref{k30} for FMEG is inevitably violated, and such a static condensation cannot exist.

We therefore focus on the remaining solutions of equation \eqref{k80}, determined by solutions of equation \eqref{k82} that satisfy \eqref{k77}.
It can be shown that such a solution of equation \eqref{k82} is unique and corresponds to the boundary condition $\psi(\ga)\to2$ as $\ga\to-\infty$.
The corresponding solutions $\ga(u)$ of equation \eqref{k80} are then determined by the relation
\disn{k83.1}{
\int\frac{d\ga}{\psi(\ga)}=u+\xi,
\nom}
where $\xi$ is an arbitrary constant, so that all $\ga(u)$ are obtained from one another by a shift of the variable $u$.
The corresponding function $\ti\ff(\ti r)$ remains finite as $\ti r\to0$ (i.e., as $u\to-\infty$).
By means of this shift in $u$ one can select a particular solution that vanishes at $\ti r=0$, which we denote by $\hat\ff(\ti r)$.
All other solutions are obtained from it by the shift $u\to u+\xi$, and accordingly from \eqref{k77a} we have
\disn{k88}{
\ti\ff(\ti r)=\hat\ff(\ti r e^\xi)+2\xi.
\nom}

\begin{figure}[htbp]\centering
\includegraphics[height=0.35\textwidth]{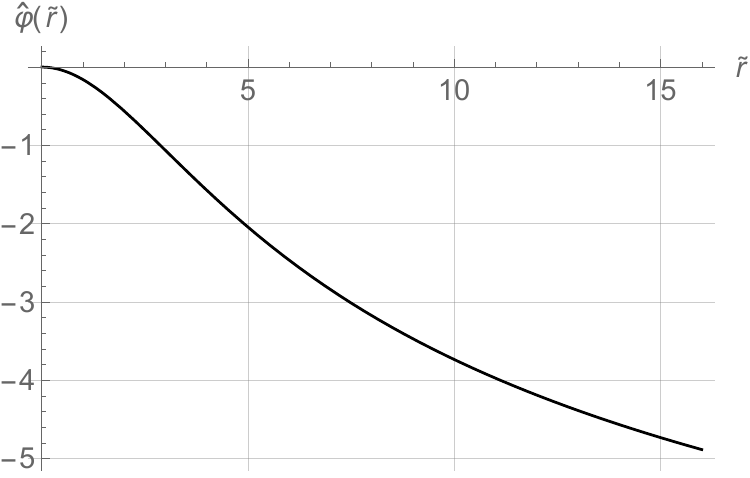}
\caption{Numerically obtained function $\hat\ff(\ti r)$.}
\label{ris1}
\end{figure}

As a result, all solutions of equation \eqref{k76} that interest us (which satisfy condition \eqref{k77} and differ from \eqref{k83}), can be written in the form
 \disn{k88a}{
\ff(r)=\ff_0-w\,\hat\ff\ls\frac{4r}{\de}\rs,
\nom}
where $\ff_0=-2w\xi$ and $\de=4e^{-\xi}\sqrt{w/C_2}>0$ are arbitrary constants introduced in place of arbitrary $C_2$ and $\xi$, and $\hat\ff(\ti r)$ is the solution of equation \eqref{k77.2} with the additional conditions $\hat\ff(\ti r)\to0$, $\ti r^2\hat\ff'(\ti r)\to0$ as $r\to0$, which turns out to be unique. Such a function $\hat\ff(\ti r)$ can be obtained numerically; its graph is shown in Fig.~\ref{ris1}.
%%%%%%%%%%%%%%%%%%%%%%%%%%%%
%%%%%%%%%%%%%%%%%%%%%%%%%%%%
Approximately at $r=\de$ one finds $\ff''(r)=0$, which means that at this distance the gravitational attraction reaches its maximum. The corresponding density distribution of FMEG \eqref{k88a} has the form
 \disn{k89}{
\rho_\ta=\rho_0 \exp{\ls\hat\ff\ls\frac{4r}{\de}\rs\rs},
\nom}
where
 \disn{k89a}{
\rho_0=\frac{4w}{\pi G\de^2}.
\nom}
The obtained spherically symmetric static condensation of FMEG, corresponding to the case where all eigenvalues of the matrix $w^{ik}$ coincide, takes the form of a sphere with a maximum density $\rho_0$ at its centre. As in the cases of the wall and string considered above, the condensation is characterised by a single parameter $\de$, which defines the characteristic size of the condensation (at $r=\de$ the density of FMEG decreases by approximately a factor of 5 compared to $\rho_0$), and the central density is related to $\de$ by similar formulae.

The asymptotic behaviour at large $r$ for the obtained spherically symmetric density profile of FMEG can be derived by perturbative analysis of \eqref{k80}, if one takes as the zeroth-order approximation the constant solution $\ga(u)=\log2$. As a result, for $r\to\infty$ one obtains the approximation
 \disn{k95}{
\rho_\ta\approx\rho_0\frac{\de^2}{8r^2}
\exp{\ls 0.6\sqrt{\frac{\de}{r}}\sin\ls\frac{\sqrt{7}}{2}\log\ls\frac{2.7r}{\de}\rs\rs\rs}.
\nom}
For a somewhat less specific formulation of the problem, an analogous result can be found in \cite{GalDyn}. Numerical verification shows that approximation \eqref{k95} works well (with a relative accuracy of about 6\%) for $r>\de$. As can be seen, up to sufficiently slow oscillations, the density of FMEG in the case of spherical condensation decreases as $1/r^2$. Such a density fall-off law in galactic haloes is well known to yield flat galactic rotation curves, so FMEG, behaving like a self-gravitating isothermal sphere, may serve as a viable model of DM.

\begin{figure}[htbp]\centering
\includegraphics[height=0.35\textwidth]{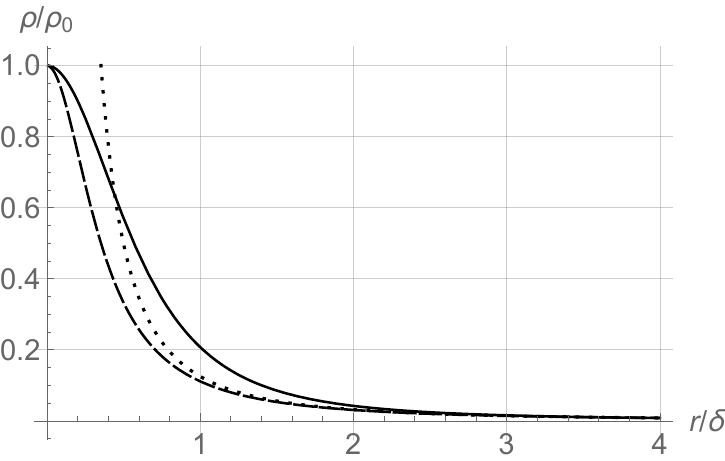}
\caption{Comparison of the isothermal \eqref{k89} (solid line), pseudo-isothermal \eqref{sf1} (dashed line), and singular isothermal \eqref{k83} (dotted line) density distribution profiles for a self-gravitating sphere.}
\label{ris2}
\end{figure}

In modelling the distribution of DM in galactic haloes, one of the frequently used options is the spherical pseudo-isothermal profile \cite{Begeman}:
 \disn{sf1}{
\rho_{\text{iso}}=\frac{\rho_0}{1+\frac{r^2}{r_0^2}}.
\nom}
This profile results from the regularization of the singular isothermal sphere \eqref{k83}. A comparison of the plots of the isothermal profile \eqref{k89} and the pseudo-isothermal profile \eqref{sf1} with identical central values and asymptotics at large $r$ (for this, $r_0=\de/\sqrt{8}$ was taken) is shown in Fig.~\ref{ris2}.
%%%%%%%%%%%%%%%%%%%%%%%%%%%%
%%%%%%%%%%%%%%%%%%%%%%%%%%%%
As can be seen, a noticeable difference between them exists only for $r/\de<2$. In the same figure, for comparison, the graph of the singular isothermal profile \eqref{k83} is presented, which simultaneously represents the $1/r^2$ asymptotic behaviour of the other two profiles at large $r$; for the isothermal profile this asymptotic behaviour works sufficiently well for $r/\de>2$.

\section{Discussion}\label{zakl}
We consider solutions of embedding gravity -- a modified theory of gravitation which can be rewritten (see \eqref{v1}-\eqref{v3}) as GR with additional fictitious matter of embedding gravity
(FMEG). We compare the properties of this matter with the observed properties of DM, attempting to explain the nature of dark matter as a purely gravitational effect within the framework of the transition from GR to embedding gravity as the fundamental theory of gravitation.

Assuming the gravitational field $g_{\m\n}$ corresponding to GR to be weak, we restrict ourselves to solutions in which the embedding function $y^a$, describing the four-dimensional surface, is represented as a small deformation \eqref{k5} of the unfolded 10-dimensional embedding of the Minkowski spacetime metric. If we consider a region of spacetime that is not too large, where the second fundamental form of the surface $b^a_{\m\n}$ varies only slightly \eqref{k13} (estimates yield the size of this region as at least several megaparsecs), then, for sufficiently small total matter density \eqref{k30}, FMEG remains in the linear regime \eqref{k14}. In this regime, the stress tensor of FMEG is proportional to its density with a constant matrix coefficient $w^{ik}$ \eqref{k34}, i.e. this fictitious matter possesses anisotropic pressure with a linear equation of state. With increasing total density, FMEG exits the linear regime, and therefore the density of FMEG cannot exceed the limiting value \eqref{k14}; in other words, fictitious matter cannot condense arbitrarily strongly. This explains why the observed effects for which the concept of DM must be invoked do not manifest themselves on very small scales (for instance, on the scale of the Solar System), where the density of ordinary matter increases.

We find possible approximately static (i.e. static in the leading approximation) configurations of FMEG,
assuming that they arise as a result of its condensation under gravitational forces
from an initially almost homogeneous distribution with positive density $\rho_\ta$.
The type of emerging configurations is determined by the structure of the eigenvalues of the matrix $w^{ik}$
in a given region of space. In regions where all eigenvalues are different
and among them there is a positive one, only condensations of the thick-wall type
with density depending on a single coordinate \eqref{k53} can occur.
Where two of the three eigenvalues coincide, having positive values,
$SO(2)$-symmetric condensations of the thick-string type
with density distribution \eqref{k69} may arise.
In both of these cases the distributions are characterised by a parameter $\de$,
associated with their transverse size, through which the central maximum density $\rho_0$
is also expressed.

When such FMEG condensations form, ordinary matter is pulled
into the corresponding gravitational potential wells. Subsequently,
as a result of the mutual attraction of its components, density fluctuations
increase, leading to the possible formation of galaxies.
Thus, the two types of FMEG condensations described above may, possibly,
be related to the two- and one-dimensional structures
that are formed by galaxies -- sheet-like tenuous walls and
long elongated filaments \cite{cosmicweb}.
However, this issue requires further investigation.

The most interesting case, studied in Sec.~\ref{sfer},
is when in a certain region of space
all three eigenvalues of the matrix $w^{ik}$ coincide, having positive values.
In this case, FMEG exhibits the properties of an isothermal ideal gas,
since it possesses ordinary (isotropic) pressure and a linear equation of state $p_\ta=w \rho_\ta$.
In such a region of space, spherically symmetric FMEG condensations of the sphere type
with density distribution \eqref{k89} may arise.
As in other cases, this distribution is characterised by the parameter $\de$,
which defines the size of the sphere and through which the maximum central density $\rho_0$
is expressed \eqref{k89a}.

After the formation of such an FMEG condensation,
ordinary matter is drawn into the resulting potential well,
which is clearly a more efficient mechanism of galaxy formation
than formation solely due to the mutual attraction of ordinary matter.
At the same time, the density profile of ordinary matter should be of the core type
(i.e. smooth at the centre) \cite{statja73}, which agrees with observations.
By contrast, when galaxies form in the absence of a preformed
spherically symmetric potential well, the profile, as shown by numerical simulations \cite{Navarro_1997},
is cusp-type (i.e. density diverges at the centre).

Outside the central region, where the density of the pulled-in ordinary matter
is significant, the total matter density is mainly determined
by the density of FMEG \eqref{k89} -- this is the region of the galactic dark halo.
If in this region $r>2\de$, then, as mentioned at the end of Sec.~\ref{sfer},
for the density profile one can use the leading asymptotic of approximation \eqref{k95},
which reduces to the dependence
 \disn{d1}{
\rho_\ta\approx\frac{\rho_0\de^2}{8r^2}=\frac{w}{2\pi Gr^2},
\nom}
where \eqref{k89a} has been used.
Let us write down the condition arising in the discussion of galactic rotation curves
(see, e.g., \cite{DEL_POPOLO_2014}):
 \disn{d2}{
\frac{v^2}{r}=\frac{G}{r^2}\int\limits_0^r dr\, 4\pi r^2\rho_\ta,
\nom}
where $v$ is the velocity of a star in the dark halo, and the integral
gives the mass enclosed within a sphere of radius $r$.
Substituting \eqref{d1}, we obtain $v(r)=\sqrt{2w}$,
i.e. at large $r$ the dependence $v(r)$ approaches a constant,
which is a characteristic property of galactic rotation curves,
observed for a large number of galaxies, see, e.g., reviews \cite{DEL_POPOLO_2014,ufn-zasov,2007.15539}.
As seen, the value of this constant $v_*$ equals $\sqrt{2w}$,
i.e. it is determined by the property of the background embedding
in the considered region of space.

If we assume that spherical condensations typically form
with the maximum possible central density for FMEG, i.e. $\rho_0=\rho_b$,
then from \eqref{k30} and \eqref{k89a} we obtain $\de=4\sqrt{w}L=2\sqrt{2}v_*L$,
i.e. the characteristic size of the condensation is proportional to $v_*$.
It is clear that the larger the size $\de$ of the potential well formed,
the more ordinary matter will be pulled into it.
Therefore, the baryonic mass $M_{bar}$ of the galaxy
should grow with increasing $v_*$, which qualitatively corresponds
to the baryonic Tully-Fisher relation, according to which $M_{bar}\sim v_*^\dz$,
where $\dz$ takes values between 3.5 and 4, see, e.g., the review \cite{ufn-zasov}.
If we assume that $M_{bar}$ is proportional to the volume of the potential well,
then $M_{bar}\sim \de^3\sim v_*^3$, which yields the value $\dz=3$,
sufficiently close to the observed one.
A more accurate derivation of the relation between $M_{bar}$ and $w$
requires further investigation.

In conclusion, one may state that if instead of GR
one uses embedding gravity to describe gravitation,
then, under certain additional assumptions on the choice of a class of solutions of the theory,
it becomes possible to explain certain effects that within GR
require the hypothesis of the existence of DM.
From this perspective, DM may be regarded as part of the
gravitational degrees of freedom, additional to the Einsteinian ones.
It is important to emphasise that these degrees of freedom are dynamical,
i.e. the fictitious matter can move independently of ordinary matter.

In the framework of further development of this approach,
it is necessary to investigate what happens to FMEG
when the expansion of the universe is taken into account.
One may expect that this will provide an understanding of the nature of the spatial regions
in which the various conditions on the eigenvalues of the matrix $w^{ik}$ \eqref{k34}
considered in this work are realised, and hence in which
the identified types of condensations arise.
This information may then be compared with the observed
large-scale structure of the distribution of galaxies.

{\bf Acknowledgments.}
This work was supported by the Ministry of Science and Higher Education of
the Russian Federation (agreement 075-15-2025-343 dated 29/04/2025 for
Saint Petersburg Leonhard Euler International Mathematical Institute at
Saint Petersburg State University).

%\bibliographystyle{..//..//my3}
%\bibliography{..//..//paston-grav-e}
%\end{document}

\end{document}